\documentclass[preprint2]{aastex}
\usepackage{graphicx}
\usepackage{epstopdf}

\shorttitle{Monitoring of Swift J1753.5-0127}
\shortauthors{Rodi et al.}

\begin{document}


\title{Long-Term Monitoring of the Black Hole Candidate \\ Swift J1753.5-0127 with \textit{INTEGRAL}/SPI}


\author{J. Rodi\altaffilmark{1,2}, E. Jourdain\altaffilmark{1,2}, and J. P. Roques\altaffilmark{1,2}}
\affil{\textsuperscript{1}Universit\'e de Toulouse; UPS-OMP; IRAP;  Toulouse, France\\
\textsuperscript{2}CNRS; IRAP; 9 Av. Colonel Roche, BP 44346, F-31028 Toulouse cedex 4, France\\}

 
\begin{abstract}
The black hole candidate Swift J1753.5-0127 went into outburst in 2005 June.  Rather than fade into quiescence as most black-hole-candidate transients do, it has remained in a low hard spectral state for most of the \( \sim 9 \) years after outburst.  The persistent emission while in a hard state is reminiscent of the black hole Cyg X-1 and the black hole candidates 1E 1740.7-2942 and GRS 1758-258.  Thus far hard X-ray/soft gamma-ray results have focused mainly on the 2005 flare, with a few additional observations during 2007.  Here, we present results from \textit{INTEGRAL}/SPI observations from  \( 2005 - 2010 \) spanning the \(22 - 650 \) keV energy range.  Spectral analysis shows a weak high-energy excess (\( \sim 2.9 \sigma \)) above a cutoff powerlaw model that is well fit by a powerlaw suggesting an additional spectral component.  Observations of Cyg X-1, 1E 1740.7-2942, and GRS 1758-258 have shown similar spectra requiring an additional high-energy component    The SPI results are compared to previously reported results for Swift J1753.5-0127 as well as observations of other sources.  
\end{abstract}




\section{Introduction}

Swift J1753.5-0127 was discovered by the \textit{Swift}/Burst Alert Telescope (BAT) on MJD 53551 (2005 June 30) \citep{burrows2005,palmer2005}.  The \(15-50\) keV flux continued to increase until \( \sim \) MJD 53560 (2005 July 9) with a peak flux of \( \sim 360 \) mCrab.  Instead of returning to a quiescent state after outburst, Swift J1753.5-0127 has maintained an average flux of \( \sim 55 \) mCrab in the \(15-50\) keV energy band.  X-ray and gamma-ray observations during the flare showed spectral features and QPO's consistent with a black hole (BH) in the low hard state \citep{miller2006,cadolle2007,zhang2007}.  Consequently, Swift J1753.5-0127 is considered a black hole candidate (BHC).  Radio observations during this time also detected the source as are expected for BH's in the low hard state, but the reported radio fluxes from \citet{cadolle2007} and \citet{soleri2010} were significantly lower than is expected from radio/X-ray correlations \citep{gallo2003,gallo2006}.  \citet{zurita2008} estimated an orbital period of \( P_{orb} = 3.23 \) h based on the period of the observed superhump while \citet{neustroev2014} used photometric and spectroscopic spectroscoptic variability to determine an orbital period of \(P_{orb} = 2.85 \) h.  A companion star has not yet been identified.  The inability to identify it in archival images suggests that the donor star is a low mass star \citep{cadolle2007,neustroev2014}.  \citet{cadolle2007,durant2009} estimate the distance to the system to be \( \sim 6-7\) kpc, which suggests that the donor is a main-sequence type K or M star.  Thus, the mass of the compact object is not well constrained.  \citet{froning2014} estimate a mass \( > 10 \textrm{ } M_{\odot} \) for a source distance of 6 kpc while \citet{neustroev2014} report the mass as \( < 5 \textrm{ } M_{\odot} \).

Most BH(C)'s do not exhibit persistent or quasi-persistent emission.  \citet{remillard2006} list 8 such sources (BH's: Cyg X-1, LMC X-1, LMC X-3, and GRS 1915+105; BHC's: 1E 1740.7-2942, 4U 1755-338, 4U 1957+115, and GRS 1758-258)  out of 20 confirmed BH's and 20 BHC's.  Of the 8 sources, Cyg X-1, 1E 1740.7-2942, and GRS 1758-258 are the only ones to spend most of the time in a hard state, as Swift J1753.5-0127 has done for most of the \( \sim 9 \) years since the flare \citep{soleri2013}.  These three sources have high-energy tails seen above \( \sim 200 \) keV.  (\citet{jourdain1994} and \citet{cadolle2006} for Cyg X-1; \citet{bouchet2009} for 1E 1740.7-2942; and \citet{pottschmidt2008} for GRS 1758-258.)  Possible mechanisms for the origin of the hard tails include Comptonization of a non-thermal population of electrons \citep{wardzinski2001}, bulk-Comptonization \citep{laurent1999}, and jet emission \citep{markoff2005}.  \textit{INTEGRAL} observations of Cyg X-1 by \citet{laurent2011} and \citet{jourdain2012a} have reported highly polarized emission \( > 200 \) keV.  The strong degree of polarization suggests the hard X-ray emission above \( \sim 200 \) keV is related to the radio jet seen from Cyg X-1 in the hard state.  

Even though Cyg X-1, 1E 1740.7-2942, GRS 1758-258, and Swift J1753.5-0127 spend most of the time in a hard state, the observed behavior of these sources are quite different.  Historically, Cyg X-1 undergoes transitions to a soft state on the timescale of a few years though since 2011 it has spent long periods in a soft state with short periods in the hard state.  Both 1E 1740.7-2942 and GRS 1758-258 are almost always observed to be in a hard state with short periods lasting weeks to a few months in a soft state before returning to a hard state.  Swift J1753.5-0127 has exhibited extremely different behavior in that it was discovered during an X-ray outburst, but did not fade back into quiescence, instead maintaining a persistent flux and remaining in a hard state for \( \sim 9 \) years.

In this paper, we present analysis of \textit{INTEGRAL}/SPI observations of Swift J1753.5-0127 beginning with the Target of Opportunity (ToO) during the flare in 2005 (MJD 53592) until 2010 March (MJD 55284) after which SPI observes the source for only short periods (less than \( \sim 20,000\) s).  The SPI results are compared to previous observations of the flare, other isolated pointed-observations of Swift J1753.5-0127 by other instruments, and to other transient (GRO J0422+32) and persistent (GRS 1758-258) BH(C)'s in the SPI bandpass.  These are the first published long-term results for Swift J1753.5-0127 at hard X-ray/soft gamma-ray energies.  

\section{Instrument and Observations}

The gamma-ray observatory the \textit{International Gamma-ray Astrophysics Laboratory} (\textit{INTEGRAL}) was launched on 2002 October 17 into an eccentric orbit with an orbital period of 3 days and an orbital inclination of \( 51.6^{\circ} \).  These parameters were selected to minimize the amount of time spent in Earth's radiation belt \citep{jensen2003}. The spectrometer \textit{INTEGRAL}/SPI consists of 19 hexagonal germanium detectors (GeD) configured in a hexagonal pattern with a tungsten coded mask and an active coincidence shield of 91 Bismuth Germinate (BGO) crystals to reduce the background rate \citep{vedrenne2003}.  The geometrical area of the GeD detectors is \(508 \textrm{ cm}^2 \).  An individual GeD is 3.2 cm in length and 69.42 mm in height.  The detectors are actively cooled to \( \sim 85\) K resulting in an energy resolution of \(2 - 8\) keV over the \(20 \textrm{ keV} - 8 \textrm{ MeV}\) energy range of the instrument \citep{roques2003}.  

In order to be able to generate images of the sky, \textit{INTEGRAL} follows a dithering pattern with \( 2^{\circ} \) off-axis pointings from the object of interest.  Each pointing lasts from \(0.5 - 1\) hour such that a complete dithering pattern is completed during an observation period.  The standard dithering pattern is a rectangular pattern with 1 on-axis pointing for the source of interest and 24 off-axis pointings.  There is a secondary dithering pattern which is hexagonal with 1 on-axis pointing and 6 off-axis pointings \citep{jensen2003}.  

\begin{figure*}[t]
  \centering
  \includegraphics[scale=0.55, angle=180,trim = 0mm 40mm 10mm 0mm, clip]{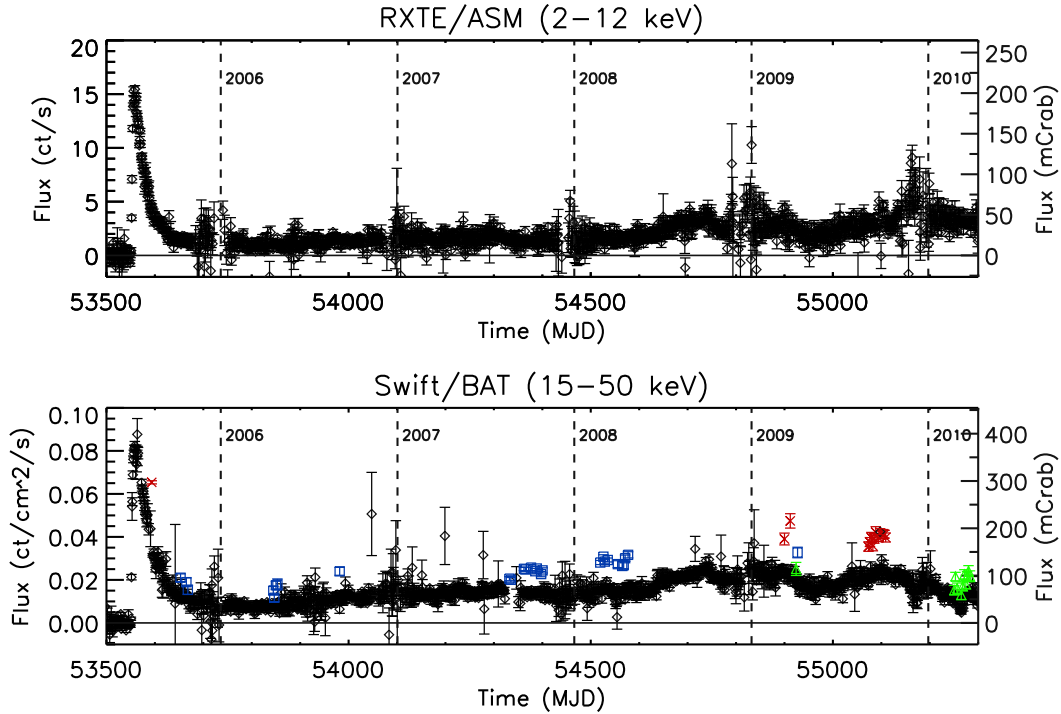}
  \caption{Top: \textit{RXTE}/ASM \(2-12\) keV lightcurve.  Periods with large errors and prolonged data gaps are times when the Sun passes nearby Swift J1753.5-0127  Bottom: \textit{Swift}/BAT lightcurve with SPI \(22-50\) keV fluxes overplotted in arbitrary flux units.  Different symbols and colors of SPI points correspond to different regions in ASM-SPI flux plot (Fig.~\ref{fig:flux}).  Legend for SPI data described in Sec.~3.2.  Vertical dashed-lines denote January 1 of the year listed to right.} \label{fig:lc}
\end{figure*}

\subsection{Observations and Data Analysis}
The only dedicated observation of Swift J1753.5-0127 by \textit{INTEGRAL} (as of revolution 1373) was as a ToO during its flare in 2005, but because of its proximity to the Galactic Plane, the source has been observed often during the mission.  Since Swift J1753.5-0127 was discovered, it has been observed within \( 12^{\circ} \) of the detector normal in at least one pointing (\( \sim 2000 \) s) in 149 \textit{INTEGRAL} revolutions.  For this work, only those revolutions with at least 10 pointings including Swift J1753.5-0127 within \( 12^{\circ} \) of the SPI pointing direction were analyzed using the SPI Data Analysis Interface (SPIDAI\footnote{Publicly available interface developed at IRAP to analyze SPI data. Available at http://sigma-2.cesr.fr/integral/spidai. See description in \citet{burke2014}}) resulting a total exposure time of 2.55 Ms over 53 revolutions.  

The flux extraction procedure in SPIDAI performs a model fit for each energy band convolving a sky model of those sources within the field of view and a background model (based on empty field observations) with the instrument response function which is then compared to the GeD count rates in the detector plane. (See \citet{jourdain2009} for more detail.)  For spectral analysis, SPI data spanning \(20-650\) keV were grouped into 50 logarithmically-spaced, energy bins.  The first two energy channels were ignored in spectral fitting because of uncertainties in the energy response \citep{jourdain2009} thus reducing the energy range to \(22-650\) keV.

\section{Results}
\subsection{Temporal Evolution}
SPI observations from 13 revolutions before the flare covering MJD \(52751 - 53490\) measured an average flux of \( 0.8 \pm 1.2 \) mCrab in the \( 22 - 50 \) keV energy band at the location of Swift J1753.5-0127.  After correcting for the positive 1 mCrab (\(  \sim 0.075 \) ct/s) bias\footnote{See more at http://heasarc.gsfc.nasa.gov/docs/xte/asm\_products.html}, the \textit{RXTE}/All-Sky Monitor (ASM) measured an average flux of \( 0.27 \pm 0.12 \) mCrab in the \(2-12\) keV band over the period of \(50088 - 53510\) (source discovered on MJD 53520) resulting in a significance of only \( 2.3 \sigma \) and is thus not statistically significant.  BAT did not detect the source, measuring an average flux of \( -0.5 \pm 0.5 \) mCrab in the \(15 - 50 \) keV band over the period of MJD \(53415 - 53510\).  Thus Swift J1753.5-0127 was not significantly detected in the \( \sim 10 \) years prior to the flare.

Fig.~\ref{fig:lc} shows the ASM \(2-12\) keV lightcurve (top panel) and the BAT \(15-50\) kev lightcurve (bottom panel) from MJD \(53500 - 55300\) (2005 May 10 to 2010 April 14) with 3-day averages.  In the bottom panel the SPI \(22-50\) keV fluxes have been overplotted in arbitrary flux units with red X's, blue squares, green triangles, and black diamonds.  The different symbols denote where the revolution falls on the ASM-SPI flux plot described in Sec.~3.2 (Fig.~\ref{fig:flux}).  The vertical dashed lines denote January 1 of the year listed to the right (e.g., the first dashed line corresponds to 2006 January 1). 

In \( \sim 10 \) days, Swift J1753.5-0127 went from undetected to a peak flux of \( \sim 360 \) mCrab in the BAT \(15-50\) keV energy band and \( \sim 220 \) mCrab in the \(2-12\) keV ASM data on MJD 53560 (2005 July 9).  The source flux then decays exponentially with a time constant of \( \tau \sim 40 \) days in the BAT data and \( \tau \sim 28 \) days in the ASM data.  It is during this decay that the \textit{INTEGRAL} ToO is performed, spanning MJD \(53593-53595\) (2005 August \(11-13\)).  The flux continues to decline until \( \sim \) MJD 53700 (2005 November 26) with a flux of \( \sim 25 \) mCrab in the BAT data and \( \sim 10 \) mCrab in the ASM data.  The flux slowly increases for \( \sim 1000\) days with 26 SPI observations during this period.  

Beginning around MJD 54636 (2008 June 19), the flux increased to \( > 100 \) mCrab in the \(15-50\) keV band for the first time since the 2005 flare \citep{krimm2008} with the \(2-12\) keV flux increasing to \( \sim 50 \) mCrab.  After this period, the Swift J1753.5-0127 lightcurve exhibits larger temporal variability in both the BAT and ASM energy ranges.  The first SPI observations after the flux increase are not until approximately MJD 54900 (2009 March 10) when Swift J1753.5-0127 was observed in four revolutions over the span of \( \sim 30 \) days.  

From about MJD \(55000-55500\) (2009 June to 2010 October), the source enters what \citet{soleri2013} refers to as a ``failed transition'' where the spectral features soften, but the source does not make a transition to a high soft state.  Swift J1753.5-0127 was in the field of view in 12 revolutions from MJD \(55073-55109\) (2009 August 30 to October 5) for observations early during the ``failed transition.''  

On roughly MJD 55190 (2009 December 25), the \(1.5-4\) keV flux seen by the \textit{MAXI}/GSC increased from \( \sim 50 \) mCrab to 100 mCrab over the span of several days \citep{negoro2009}.  During this period, the \(4-20\) keV \textit{MAXI} flux remained relatively unchanged while \(15-50\) keV BAT flux showed a gradual decrease, suggesting a hard-to-soft state transition that lasted for \( \sim 200 \) days before the \(15-50\) keV flux recovered.  The final group of SPI observations (10 revolutions) occurred after this hard-to-soft transition and covers MJD \(55249-55284\) (2010 February 22 to March 29).  

\newcounter{NameOfTheNewCounter}
\stepcounter{NameOfTheNewCounter}
\newcounter{NameOfTheNewCounter0}
\stepcounter{NameOfTheNewCounter0}
\subsection{Spectral Evolution}
\begin{figure}[t]
  \centering
  \includegraphics[scale=0.9,angle=180,trim = 160mm 130mm 20mm 0mm, clip]{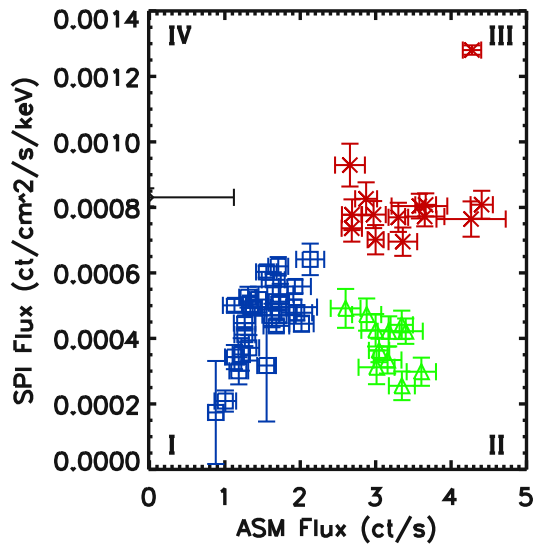}
  \caption{SPI \(22-50\) keV flux vs 3-day ASM \(2-12\) keV flux corresponding to the time of the SPI observation.} \label{fig:flux}
\end{figure}

The \(22-50 \) keV SPI flux derived from SPIDAI and the \( 2-12 \) keV ASM average flux corresponding to the time of the revolution have been plotted in Fig.~\ref{fig:flux}.  The fluxes separate into four distinct regions above and below \( 0.00065 \) ct/cm\(^2\)/s/keV in the SPI flux and above and below \( 2.5 \) ct/s in the ASM flux. Region \Roman{NameOfTheNewCounter} corresponds to the bottom left quadrant (both fluxes low), Region \setcounter{NameOfTheNewCounter}{2}\Roman{NameOfTheNewCounter} to the bottom right quadrant (SPI flux low and ASM flux high), Region \setcounter{NameOfTheNewCounter}{3}\Roman{NameOfTheNewCounter} to the top right quadrant (both fluxes high), and Region \setcounter{NameOfTheNewCounter}{4}\Roman{NameOfTheNewCounter} to the top left quadrant (SPI flux high and ASM flux low).  In Fig.~\ref{fig:lc} and~\ref{fig:flux}, revolutions in Region \Roman{NameOfTheNewCounter0} are marked by blue squares,\stepcounter{NameOfTheNewCounter0} Region \Roman{NameOfTheNewCounter0} by green triangles, \stepcounter{NameOfTheNewCounter0} Region \Roman{NameOfTheNewCounter0} by red X's, and Region \stepcounter{NameOfTheNewCounter0}\Roman{NameOfTheNewCounter0} by black diamonds.  Region \Roman{NameOfTheNewCounter} contains only one point, which has a negative ASM flux with a large error.  This point corresponds to Revolution 848 in the SPI data, and the ASM flux is based on a single measurement with a 90 s exposure.  Revolution 848 has been analyzed in Region \setcounter{NameOfTheNewCounter}{3}\Roman{NameOfTheNewCounter} as its SPI flux is consistent with that region and because revolutions \(840-845\), 847, and \(849-852\) are also in Region \Roman{NameOfTheNewCounter}.  This leaves Region \setcounter{NameOfTheNewCounter}{4}\Roman{NameOfTheNewCounter} empty.  

In Region \setcounter{NameOfTheNewCounter}{1}\Roman{NameOfTheNewCounter}, the SPI and ASM fluxes show a roughly linear correlation while in Region \setcounter{NameOfTheNewCounter}{2}\Roman{NameOfTheNewCounter} the SPI and ASM fluxes show an anticorrelation.  Region \setcounter{NameOfTheNewCounter}{3}\Roman{NameOfTheNewCounter} presents a different relationship between the SPI and ASM fluxes.  When excluding the flare observation (the highest SPI flux), the SPI flux is roughly constant while the ASM flux varies by a factor of \( \sim 2 \).  

The individual revolution spectra in a region were summed together to create a summed spectrum with better statistics.  The total exposure is 1.69 Ms for Region \setcounter{NameOfTheNewCounter}{1}\Roman{NameOfTheNewCounter}, 0.31 Ms for Region \setcounter{NameOfTheNewCounter}{2}\Roman{NameOfTheNewCounter}, 0.55 Ms for Region \setcounter{NameOfTheNewCounter}{3}\Roman{NameOfTheNewCounter} with 0.14 Ms of the Region \Roman{NameOfTheNewCounter} exposure during the flare.  The summed spectrum for each region was fit in XSPEC with a powerlaw model and a cutoff powerlaw model. 

\begin{figure*}[t]
  \centering
  \includegraphics[scale=0.675,angle=180,trim = 10mm 20mm 10mm 15mm, clip]{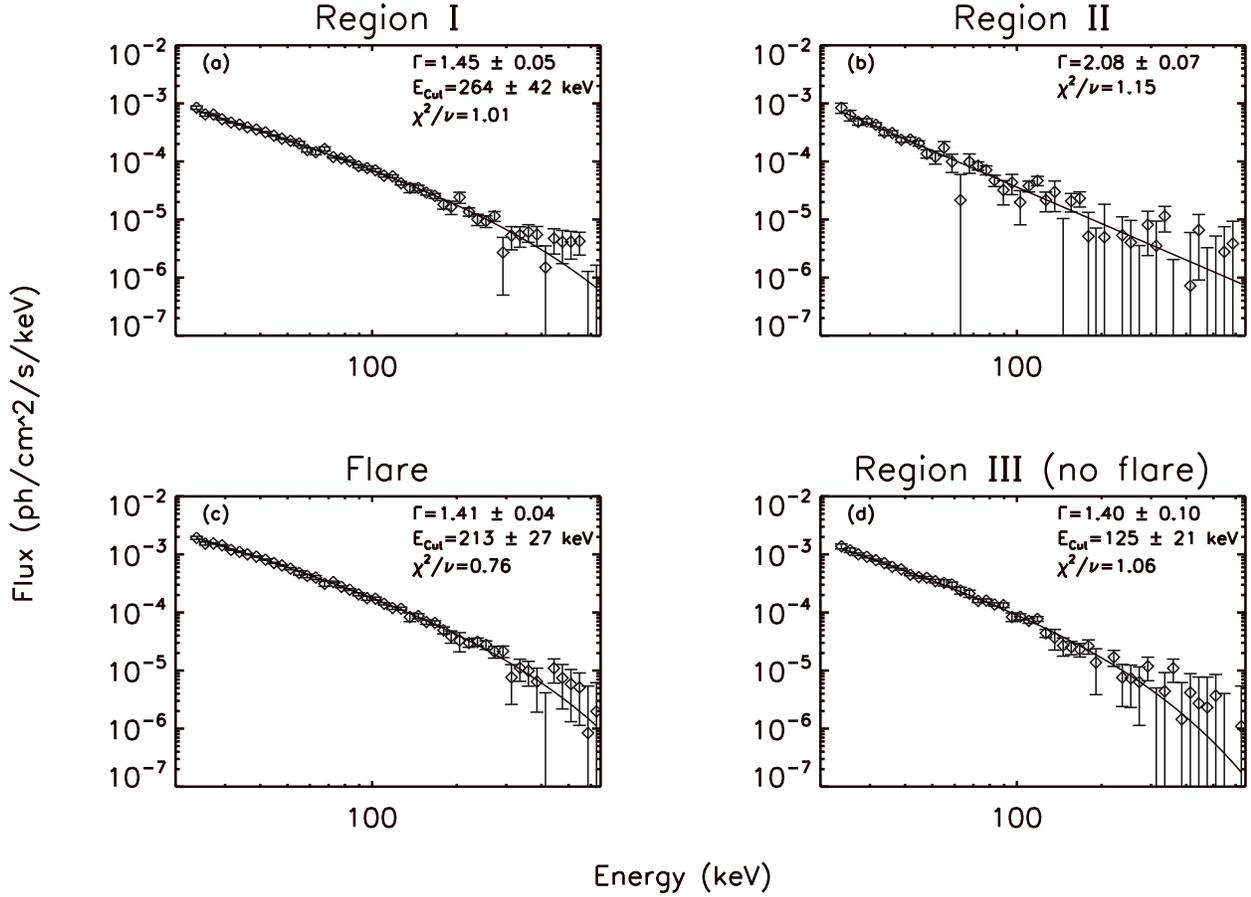}
  \caption{(a) Region \setcounter{NameOfTheNewCounter}{1}\Roman{NameOfTheNewCounter} spectra fit to cutoff powerlaw model (b) Region \setcounter{NameOfTheNewCounter}{2}\Roman{NameOfTheNewCounter} spectra fit to powerlaw model (c) Spectra of \textit{INTEGRAL} ToO fit to cutoff powerlaw model (d) Region \setcounter{NameOfTheNewCounter}{3}\Roman{NameOfTheNewCounter} (excluding flare) fit to cutoff powerlaw model } \label{fig:spectra}
\end{figure*}

Region \setcounter{NameOfTheNewCounter}{1}\Roman{NameOfTheNewCounter} consists primarily of observations during the tail of the flare and the slow flux increase.  For the powerlaw model, Region \Roman{NameOfTheNewCounter} has a best-fit spectral index of \( \Gamma = 1.75  \pm 0.01  \) (\(1 \sigma\) errors) and has a \( \chi^2 / \nu =2.10  \) (\(\nu = 46\)).  A cutoff powerlaw model better fits the data resulting in \( \chi^2 / \nu = 1.01 \) (\( \nu = 45\)) with  \( \Gamma = 1.45  \pm 0.05  \) and \( \textrm{E}_{\textrm{cut}} = 264 \pm 42 \) keV (Fig.~\ref{fig:spectra}a).  The Region \setcounter{NameOfTheNewCounter}{2}\Roman{NameOfTheNewCounter} spectrum is predominately from observations after the BAT flux decrease but the \(4-20\) \textit{MAXI} flux remained relatively constant.  Region \Roman{NameOfTheNewCounter} is well fit by a powerlaw model with \( \Gamma = 2.08 \pm 0.07 \) and has \( \chi^2 / \nu = 1.15 \) (\( \nu = 46\)) (Fig.~\ref{fig:spectra}b).  When fit to a cutoff powerlaw model, the best-fit parameters are \( \Gamma = 1.95 \pm 0.21 \) and \( \textrm{E}_{\textrm{cut}} = 546 \pm 711\) keV with \( \chi^2 / \nu =1.17  \) (\(\nu = 45\)).  The cutoff energy is not constrained and thus is not required.  

Region \setcounter{NameOfTheNewCounter}{3}\Roman{NameOfTheNewCounter} contains the ToO observation during the flare and observations shortly after the ``failed transition'' began.  The SPI flux from the ToO is significantly higher than the rest of the region, and consequently that observation will be analyzed independently from the other Region \Roman{NameOfTheNewCounter} observations.  When the ToO observation is fit with a powerlaw model, the best fit has  \( \Gamma = 1.78 \pm 0.01 \) with \( \chi^2 / \nu = 2.74 \) (\( \nu = 46 \)).  The data are better fit by a cutoff powerlaw model with \( \Gamma = 1.41 \pm 0.04 \) and \( \textrm{E}_{\textrm{cut}} = 213 \pm 27 \) keV (\( \chi^2 / \nu = 0.76 \) and \( \nu = 45\)) (Fig.~\ref{fig:spectra}c).  When the rest of Region \Roman{NameOfTheNewCounter} is fit with a powerlaw model, the best-fit photon index is \( \Gamma = 1.96 \pm 0.03 \) with \( \chi^2 / \nu = 1.69 \) (\( \nu = 46 \)).  The cutoff powerlaw model is best fit by \( \Gamma = 1.40 \pm 0.10 \) and \( \textrm{E}_{\textrm{cut}} = 125 \pm 21 \) keV (\( \chi^2 / \nu = 1.06 \) and \( \nu = 45\)) (Fig.~\ref{fig:spectra}d).  The spectral index is similar to Region \setcounter{NameOfTheNewCounter}{1}\Roman{NameOfTheNewCounter} and the flare, but the cutoff energy has decreased.  

Because of similar best-fit parameters, Regions \setcounter{NameOfTheNewCounter}{1}\Roman{NameOfTheNewCounter} and \setcounter{NameOfTheNewCounter}{3}\Roman{NameOfTheNewCounter} (including the flare) were combined for an average cutoff powerlaw spectrum.  This average spectrum has best fit parameters of \( \Gamma = 1.47 \pm 0.04 \) and \( \textrm{E}_{\textrm{cut}} = 234 \pm 26 \) keV with \( \chi^2 / \nu = 1.01 \) (\( \nu = 45\)) (Fig.~\ref{fig:avgspec}) with significant flux above \(  300 \textrm{ keV } ( > 6 \sigma ) \).  A weak excess above the cutoff powerlaw (\( \sim 2.9 \sigma\)) was present between 400 and 600 keV.  To try to fit the excess, a powerlaw was added to the cutoff powerlaw model.  The best-fit parameters are \( \Gamma_1 = 1.12 \pm 0.82 \textrm{, E}_{\textrm{cut}} = 102 \pm 82 \textrm{ keV, and } \Gamma_2 = 1.75 \pm 0.42 \textrm{ with } \chi^2 / \nu = 0.97 \) (\( \nu = 43\)).  The \( \chi^2 / \nu \) is slightly smaller compared to the cutoff powerlaw model, but the fit parameters are poorly constrained.  

\section{Discussion}
\subsection{Comparison with Other Swift J1753.5-0127 Observations}
The first SPI observations of Swift J1753.5-0127 during outburst are well fit to a cutoff powerlaw with a low spectral index of \( \Gamma = 1.41 \) and a cutoff energy \( \sim 200 \) keV.  These parameters are consistent with a BH in a hard state \citep{remillard2006}.  \citet{cadolle2007}  fit \textit{INTEGRAL} and \textit{RXTE} observations from the ToO over the energy range from 3 keV to 1 MeV.  The initial model used was the Comptonization model by \citet{titarchuk94} convolved with an absorption model resulting in a seed photon temperature of \( kT_0 = 0.51 \pm 0.08 \textrm{ keV, an electron temperature of } kT_e = 88 \pm 14 \textrm{ keV, an optical depth of } \tau = 0.67 \pm 0.14 \) with large residuals \( < 20 \) keV.  When incorporating a reflection component to better fit the data \( < 20 \) keV and performing the fit using only PCA, IBIS/ISGRI, and SPI data, an acceptable \( \chi^2 / \nu \) was found.  The best-fit parameters were a seed photon temperature of \( kT_0 = 0.54_{-0.07}^{+0.04} \textrm{ keV, an electron temperature of } kT_e = 150 \pm 26 \textrm{ keV, an optical depth of } \tau = 1.06 \pm 0.02 \textrm{, and a reflection fraction of } \Omega/2\pi = 0.32 \pm 0.03 \).  The addition of the reflection component significantly increased both the electon temperature and the optical depth.

Because the residuals for the compTT fit in \citet{cadolle2007} were below the SPI energy range, an initial fit was performed without the reflection component using the seed photon temperature (0.51 keV) from \citet{cadolle2007}.  The best-fit parameters for this model are an electron temperature of \(  kT_e = 94 \pm 24 \textrm{ keV and an optical depth of } \tau = 0.74 \pm 0.22 \textrm{ with } \chi^2 / \nu = 0.86 \) (\( \nu = 45 \)).  When the reflection component is included and the photon temperature (0.54 keV) and reflection fraction (0.32) are fixed to the values from \citet{cadolle2007}, the best-fit parameters are \(  kT_e = 98 \pm 27 \textrm{ keV and an optical depth of } \tau = 0.79 \pm 0.25 \textrm{ with } \chi^2 / \nu = 0.74 \) (\( \nu = 45 \)).  Including the reflection component to the model improves the fit to the data compared to the compTT model alone while the parameters remain consistent.  When fitting without the reflection component, the electron temperatures and optical depths are consistent with the corresponding fit in \citet{cadolle2007}.  When the reflection component is added to the fit, the electron temperatures are inconsistent and the optical depths are marginally consistent to the corresponding model in \citet{cadolle2007}.  

There have been few published spectral results at hard X-ray/soft gamma-ray energies of Swift J1753.5-0127 after the flare and none after 2007 observations by \textit{RXTE} \citep{durant2009} and \textit{Suzaku} \citep{reynolds2010}.  The \textit{RXTE} \citep{durant2009} observations took place on MJD 54262 and 54264 (2007 June 11 and 13) using both PCA and HEXTE.  The data was fit over the \(2-50\) keV energy range with a powerlaw \( \Gamma = 1.548 \pm 0.005 \).  The nearest SPI observations are \( \sim 80 \) days after thus a (quasi-)simultaneous comparison cannot be made.  These observations occur during Region \setcounter{NameOfTheNewCounter}{1}\Roman{NameOfTheNewCounter} so a reasonable comparison can be made of the Region \Roman{NameOfTheNewCounter} average spectrum in the overlapping energy range.  In the \(22-50\) keV range, the SPI data are best fit to a powerlaw with \( \Gamma = 1.59 \pm 0.05 \) with \( \chi^2 / \nu = 0.91 \) (\( \nu = 10\)) and so are consistent with these observations.

The \textit{Suzaku} observations in \citet{reynolds2010} allow for a more direct comparison with data covering MJD \(54362-54366\) (2007 September \(19 - 23\)).  These days overlap with \textit{INTEGRAL} revolutions 602 and 603 which cover MJD \(54361-54366\) (2007 September \(18 - 23\)).  \citet{reynolds2010} report no need for a cutoff model when fitting the data over the \(2-150\) keV energy range.  The data are best fit to a powerlaw model with \( \Gamma = 1.619 \pm 0.003 \textrm{ or } \Gamma = 1.608 \pm 0.003\) depending on the \( N_H \) value used in their analysis.  When the SPI data  for these revolutions are fit over the \(22-650\) keV energy range to a powerlaw the best-fit spectral index is significantly steeper with \( \Gamma = 1.75 \pm 0.04 \textrm{ with } \chi^2 / \nu = 1.19 \) (\( \nu = 46\)).  When the SPI data are fit to a cutoff powerlaw model, the best-fit parameters are \( \Gamma = 1.35 \pm 0.13 \textrm{ and } \textrm{E}_{\textrm{cut}} = 198 \pm 68 \) keV with  \( \chi^2 / \nu = 0.97 \textrm{ } (\nu = 45 ) \).  The lower \( \chi^2 / \nu \) indicates the presence of curvature in the spectrum even if the cutoff energy is not well constrained.  However, when the  energy range of the SPI data is reduced to \(22-150\) keV, the best-fit spectral index is \( \Gamma = 1.64 \pm 0.05 \) with \(\chi^2 / \nu = 0.76 \) (\( \nu = 25 \)), consistent with the spectral indexes from the \textit{Suzaku} results.  

\subsection{Comparison with other BH(C)'s}
\subsubsection{Comparison of the Flare with GRO J0422+32}
The lightcurve of Swift J1753.5-0127 during the flare (Fig.~\ref{fig:lc}) displays the typical characteristics of an X-ray nova with a fast rise with a slow decline \citep{tanaka1996}.  Similar behavior was seen in the black hole X-ray novae A 0620-003 \citep{elvis1975,kaluzienski1977}, GS 2000+251 \citep{tsunemi1989}, GRS 1124-684 \citep{kitamoto1992,ebisawa1994}, and GRO J0422+32 \citep{harrison1994}.  The lightcurve of Swift J1753.5-0127 differs from these sources in that instead of fading into quiescence, Swift J1753.5-0127 has maintained a flux between \( \sim 50-100 \) mCrab in the \(15-50\) keV energy band for \( \sim 9 \) years after the initial outburst.  

Swift J1753.5-0127 has some other similarities to GRO J0422+32 as both sources are located at high galactic latitude \citep{cadolle2007,zurita2008}, have short orbital periods (\( < 3.5 \) hours for Swift J1753.5-0127 and 5.1 hours \citep{filippenko1995} for GRO J0432+22), and have been observed in the hard state during outburst.  GRO J0422+32 was detected up to 600 keV with a hard tail by \textit{GRANAT}/SIGMA \citep{roques1994}.  \citet{vandijk1995} combined the SIGMA data with COMPTEL data from the \textit{Compton Gamma-Ray Observatory} for a spectrum spanning 35 keV to 30 MeV.  This joint spectrum was fit with a compTT model and gave best-fit parameters with an electron temperature of \(  kT_e = 100 \pm 4 \textrm{ keV and an optical depth of } \tau = 1.04 \pm 0.05 \) with a high-energy deviation \( > 300 \) keV.  The electron temperature for GRO J0422+32 is similar to the results from this work with compTT models with and without the reflection component (98 and 94 keV, respectively).  The optical depth from \citet{vandijk1995} is larger than the optical depths from compTT fit to Swift J1753.5-0127 (0.74 and 0.79 with and without a reflection component).

\begin{figure}[h]
  \centering
  \includegraphics[scale=0.75,angle=180,trim = 140mm 20mm 20mm 90mm, clip]{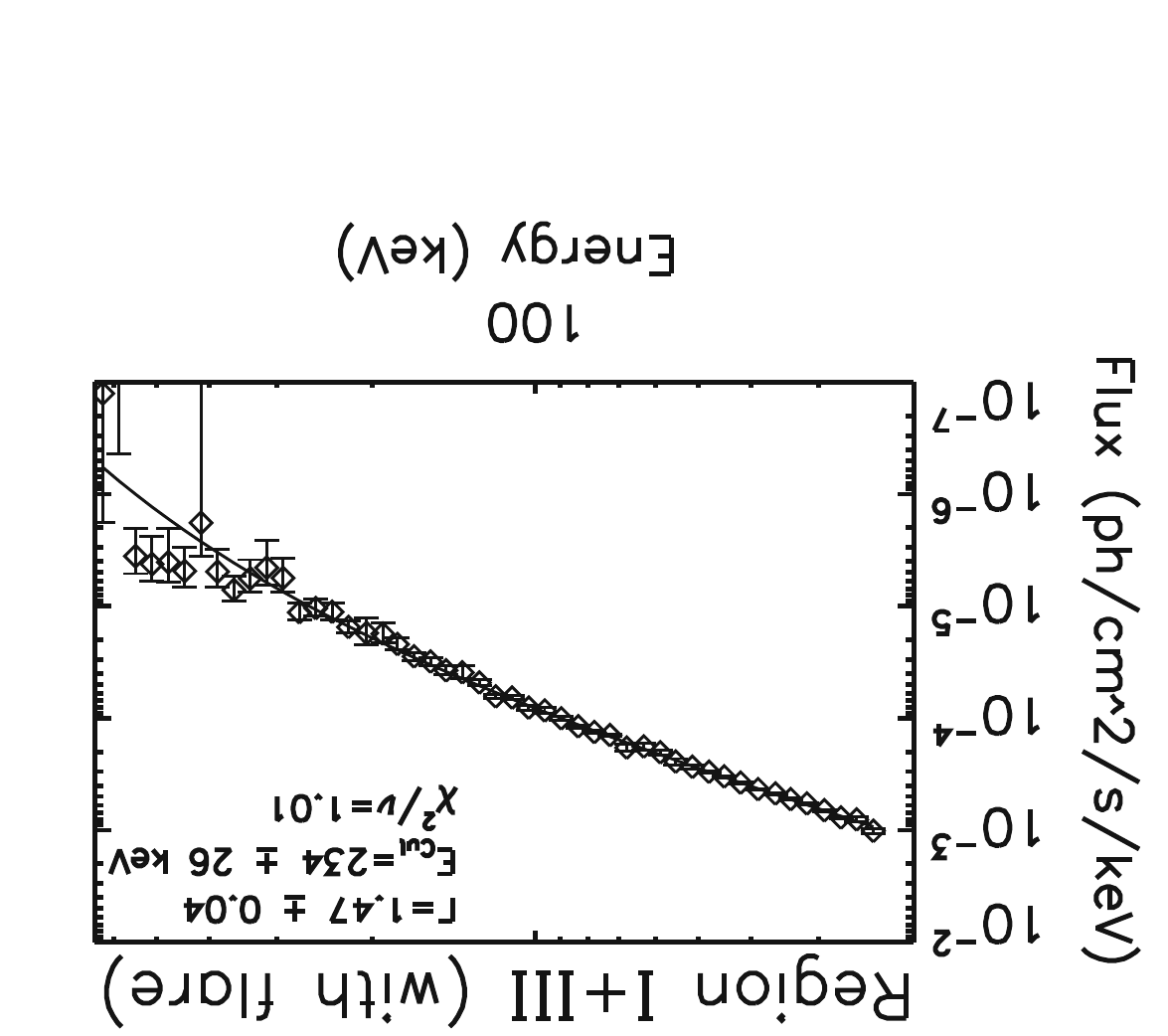}
  \vspace{-20mm}
  \caption{Summed spectrum of Regions \setcounter{NameOfTheNewCounter}{1}\Roman{NameOfTheNewCounter} and \setcounter{NameOfTheNewCounter}{3}\Roman{NameOfTheNewCounter} (including the flare) } \label{fig:avgspec}
\end{figure}

\subsubsection{Comparison of the Persistent Emission with GRS 1758-258}
As has been reported by \citet{soleri2013}, Swift J1753.5-0127 remained in a spectral state consistent with a BH in a low hard state for years after its initial outburst.  The presence of persistent emission from BH(C) transients is uncommon as most return to quiescence within weeks to months after outburst.  The flare, Region \setcounter{NameOfTheNewCounter}{1}\Roman{NameOfTheNewCounter} and Region \setcounter{NameOfTheNewCounter}{3}\Roman{NameOfTheNewCounter} show thermal Comptonized spectra typical of BH(C)'s in a low hard state and are well fit by a cutoff powerlaw model with a cutoff energy \( > 100 \) keV.  BH(C)'s are often compared to Cyg X-1 as it is the most studied BH, and Swift J1753.5-0127 has displayed behavior similar to Cyg X-1 in its timing behavior \citep{soleri2013} and in its broad band spectral behavior.  But the Galactic Center source GRS 1758-258 perhaps provides a closer analogue than Cyg X-1 as both Swift J1753.5-0127 and GRS 1758-258 are thought to have a low mass companion star and to undergo mass transfer by Roche lobe overflow \citep{neustroev2014,rothstein2002} while Cyg X-1 has a high mass companion transferring mass by stellar winds \citep{lamers1976}.  Swift J1753.5-0127 and GRS 1758-258 also show radio fluxes below the expected values based on observed radio/X-ray correlations \citep{soleri2010}.  

As part of \textit{INTEGRAL}'s 2003 and 2004 observation schedule, the satellite spent an extensive amount of time monitoring the Galactic Plane and the Galactic Center allowing for numerous observations of GRS 1758-258.  \citet{pottschmidt2006} grouped \textit{INTEGRAL}/ISGRI and SPI data into four ``epochs,'' roughly two months long each.  Biweekly PCA observations were also included for an energy range of 3-500 keV.  Epoch 1 took place during a ``dim state'' where the hard X-ray flux decreased by over an order of magnitude compared to the average hard X-ray flux.  The ``dim state'' is characterized by the blackbody thermal spectrum of a soft state (\( \Gamma = 2.29 \)) but the source does not transition to a high flux at soft X-rays and is similar to a ``failed transition'' \citep{pottschmidt2006}.  Based on flux levels at soft and hard X-rays, Epoch 1 is analogous to Region \setcounter{NameOfTheNewCounter}{1}\Roman{NameOfTheNewCounter} where both ASM and SPI fluxes are low, but Region \Roman{NameOfTheNewCounter} shows a Comptonization spectrum with a high energy cutoff \( \sim 250 \) keV while the ``dim state'' shows a soft spectrum.  In terms of spectral shape, Epoch 1 is more similar to Region \setcounter{NameOfTheNewCounter}{2}\Roman{NameOfTheNewCounter} (low SPI flux and high ASM flux) for Swift J1753.5-0127.  While Region \setcounter{NameOfTheNewCounter}{2}\Roman{NameOfTheNewCounter} does not display the dramatic flux decrease during the ``failed transition'' that GRS 1758-258 does in the ``dim state,'' though both periods show similar blackbody spectra with \( \Gamma > \sim 2.1 \).  

Epochs 2, 3, and 4 are well fit by a cutoff powerlaw model with spectral indexes ranging from \( \Gamma = 1.54 - 1.69 \) and cutoff energies from \( \textrm{E}_{\textrm{cut}} = 136 - 246 \) keV. The Swift J1753.5-0127 Comptonization spectra from Regions \setcounter{NameOfTheNewCounter}{1}\Roman{NameOfTheNewCounter} and \setcounter{NameOfTheNewCounter}{3}\Roman{NameOfTheNewCounter} show harder powerlaw indexes than those from GRS 1758-258, but the cutoff energies fall within the same range as expected for BH(C)'s.  When extending the GRS 1758-258 observations to include 11 epochs, \citet{pottschmidt2008} found  epochs 2 through 9 in a hard state.  The combined spectrum for epochs \( 2-9 \) showed a statistically significant hard tail extending to 800 keV.  

\citet{pottschmidt2008} finds that a compTT + powerlaw model better describes the data than a cutoff powerlaw + powerlaw model, giving best-fit parameters \( kT_e = 41 \textrm{ keV, an optical depth of } \tau = 1.4 \textrm{ a spectral index of } \Gamma = 1.4 \).  When the combined Swift J1753.5-0127 spectrum (Fig.~\ref{fig:avgspec}) is fit to a compTT + powerlaw model, the best-fit parameters are  \( kT_e = 39 \pm 8 \textrm{ keV, an optical depth of } \tau = 1.50 \pm 0.25 \textrm{, and a spectral index of } \Gamma = 1.31 \pm 0.89 \) with \( \chi^2 / \nu = 0.93 \) (\(\nu = 41 \)).  Compared to GRS 1758-258, the parameters for Swift J1753.5-0127 are similar, even if poorly constrained, as is the case for the spectral index.  Also, this model better describes the data than the cutoff powerlaw + powerlaw model (\( \chi^2 / \nu = 0.97\)).  As a comparison, for 1E 1740.7-2942 \citet{bouchet2009} found a spectral index of \( \Gamma = 1.9 \).  For Cyg X-1, \citet{jourdain2012b} showed that a cutoff component is required to fit the high-energy component, which can be described by a cutoff powerlaw model with \( \Gamma = 1.6 \) and a cutoff energy of 700 keV \citep{jourdain2012a}.

Swift J1753.5-0127 displayed spectral behavior similar to the BHC GRS 1758-258 in both a hard spectral state and in a ``failed transition'' state as well as having a high-energy excess, though weak in the case of Swift J1753.5-0127.  The excesses for Swift J1753.5-0127 and GRS 1758-258 are well described by hard powerlaw indexes \( \Gamma = 1.3-1.4\) while 1E 1740.7-2942 is well fit by a steeper index, and Cyg X-1 requires a cutoff model for the high-energy component.  

\section{Conclusion}
In this work, we presented the first published hard X-ray/soft gamma-ray observations of Swift J1753.5-0127 since 2007 observations with RXTE \citep{durant2009} and \textit{Suzaku} \citep{reynolds2010}.  The \textit{INTEGRAL}/SPI observations span the decreasing portion of the flare in 2005 (MJD 53592) until 2010 March (MJD 55284).  Because of SPI's sensitivity at high energies, we have been able to detect the high-energy cutoff of Swift J1753.5-0127 not seen since the flare.  The source remained in a persistent, low-hard flux state for a majority of the SPI observations, undergoing a ``failed transition'' \citep{soleri2013} beginning in 2009 June (\( \sim \) MJD 54983).

Analysis of SPI data for 53 \textit{INTEGRAL} revolutions when combined with \textit{RXTE}/ASM data revealed three distinct flux regions after the initial outburst.  For each region, the summed spectrum was fit to a powerlaw model and a cutoff powerlaw model over the \(22-650\) keV energy range.  Region \setcounter{NameOfTheNewCounter}{1}\Roman{NameOfTheNewCounter} (when both SPI and ASM fluxes were low) was well fit by a cutoff powerlaw model with a spectral index of 1.45 and a cutoff energy of 265 keV.  Region \setcounter{NameOfTheNewCounter}{2}\Roman{NameOfTheNewCounter} (when the SPI flux was low and the ASM flux was high) was well fit by a powerlaw model with a spectral index of 2.08.  Region \setcounter{NameOfTheNewCounter}{3}\Roman{NameOfTheNewCounter} (when the SPI flux was roughly constant and the ASM flux varied by \( \sim 2\)) was well fit by a cutoff powerlaw model with a spectral index of 1.40 and a cutoff energy of 125 keV when excluding the ToO observation from the fit.  When the ToO was fit alone with a cutoff powerlaw model, the best-fit parameters were a spectral index of 1.41 and a cutoff energy of 213 keV.  The data that were well-fit by a cutoff powerlaw were then combined for an average cutoff powerlaw spectrum with parameters \( \Gamma = 1.47 \textrm{ and } \textrm{E}_{\textrm{cut}} = 234 \) keV and having a weak excess (\(2.9 \sigma \)) from \(400-600\) keV.  This spectrum was better fit by a compTT + powerlaw model than a cutoff powerlaw + powerlaw model, resulting in best-fit parameters \( kT_e = 39 \textrm{ keV, an optical depth of } \tau = 1.50 \textrm{, and a spectral index of } \Gamma = 1.31 \).

In conclusion, Swift J1753.5-0127 presents the unusual case of a BHC initially behaving like a typical X-ray nova until failing to fade into quiescence.  The source showed some temporal similarities to GRO J0422+32 during the flare with a fast rise time with an slow decay as well as some spectral similarities as the electron temperatures and optical depths of both sources are similar.  The continued emission for Swift J1753.5-0127 while in a hard spectral state puts the source in a small group of BH(C)'s consisting of Cyg X-1, 1E 1740.7-2942, and GRS 1758-258.  These four sources have (quasi-)persistent emission and remain in a hard state most of the time.  Above \( \sim 200 \) keV, the spectra of Cyg X-1, 1E 1740.7-2942, and GRS 1758-258 show an additional spectral component.  Swift J1753.5-0127 shows a weak excess above \( \sim 200 \) keV but is not as bright as Cyg X-1 and lacks the exposure time of 1E 1740.7-2942 and GRS 1758-258 (\( \sim 8 \) Ms of \textit{INTEGRAL} observations).  

  For Cyg X-1, polarization measurements suggest that the hard tail is associated with a jet.  As Cyg X-1, 1E 1740.7-2942, and GRS 1758-258 are all microquasars \citep{stirling2001,mirabel1992,rodriguez1992}, the hard tails for 1E 1740.7-2942 and GRS 1758-258 could also be related to jets.  Interestingly, \citet{soleri2010} report 1E 1740.7-2942, GRS 1758-258, and Swift J1753.5-0127 as having a similar radio/X-ray relationship (with Cyg X-1 not reported).  Assuming the persistent emission continues, Swift J1753.5-0.127, as well as 1E 1740.7-2942 and GRS 1758-258, are a candidates to look for polarization above the thermal Comptonization component with a future mission that has higher polarization sensitivity.

\section*{Acknowledgments}  The \textit{INTEGRAL} SPI project has been completed under the responsibility and leadership of CNES.  We are grateful to ASI, CEA, CNES, DLR, ESA, INTA, NASA and OSTC for support.

\end{document}